\documentclass[%
 aip,
 jmp,%
 amsmath,amssymb,
preprint
]{revtex4-1}

\usepackage{graphicx}
\usepackage{dcolumn}
\usepackage{bm}
\renewcommand{\vec}[1]{\boldsymbol{#1}}

\usepackage{array}
\usepackage{color} 

\begin{document}

\preprint{}

\title[Ultrafast demagnetization in FePt]{Speed limit of FePt spin dynamics on femtosecond timescales}

\author{J. Mendil}
 \affiliation{I. Physikalisches Institut, Universit\"at G\"ottingen, Friedrich-Hund Platz 1, 37077 \mbox{G\"ottingen, Germany}}%
\author{P. C. Nieves}
  \affiliation{Instituto de Ciencia de Materiales de Madrid, CSIC, Cantoblanco, \mbox{28049 Madrid, Spain}}%
\author{O. Chubykalo-Fesenko}
	  \affiliation{Instituto de Ciencia de Materiales de Madrid, CSIC, Cantoblanco, \mbox{28049 Madrid, Spain}}%
\author{J. Walowski}
 \affiliation{I. Physikalisches Institut, Universit\"at G\"ottingen, Friedrich-Hund Platz 1, 37077 \mbox{G\"ottingen, Germany}}%
\author{M. M\"unzenberg}
 \email{corresponding author: mmuenze@gwdg.de}
 \affiliation{I. Physikalisches Institut, Universit\"at G\"ottingen, Friedrich-Hund Platz 1, 37077 \mbox{G\"ottingen, Germany}}%
\author{T. Santos}
\author{S. Pisana}
\affiliation{San Jose Research Center, HGST, a Western Digital Company, 3403 Yerba Buena Rd., San Jose, \mbox{California 95135, USA}}
\date{\today}

\begin{abstract}
Magnetization manipulation is becoming an indispensable tool for both basic and applied research. Theory predicts two types of ultrafast demagnetization dynamics classified as type I and type II. In type II materials, a second slower process takes place after the initial fast drop of magnetization. In this letter we investigate this behavior for FePt recording materials with perpendicular anisotropy. The magnetization dynamics have been simulated using a thermal micromagnetic model based on the Landau-Lifshitz-Bloch equation. We identify a transition to type II behavior and relate it to the electron temperatures reached by the laser heating. This slowing down is a fundamental limit to reconding speeds in heat assisted reversal.

\end{abstract}

\pacs{Valid PACS appear here}
\maketitle

The FePt Ll$_0$ alloy represents the most important material for novel concepts in magnetic recording due to its high magnetic anisotropy, which ensures long-time thermal stability of nanometer sized bits\cite{Moser}. Thin films of FePt with perpendicular anisotropy and small grain sizes are the most promising candidate for heat-assisted magnetic recording, which could reach storage densities beyond 1 Tb/inch$^2$. Patterning continuous FePt into individual bits\cite{Stipe} can in principle extend recording densities to 100 Tb/inch$^2$. The ultimate magnetic recording applications will also require faster bit switching. However, non-deterministic fractioning in ultrafast magnetization reversal can limit the switching speed in recording schemes, and thus has inspired fundamental research for nearly two decades\cite{Back}. Recently a new concept of ultrafast all-optical magnetic recording with an unprecedented switching timescale below 1 ps was suggested\cite{Stanciu}. This opened new possibilities to reduce the speed limit established by the spin-orbit coupling timescale to that governed by the much stronger exchange interaction. Here we show that in FePt fractioning limits the ultimate switching speed through critical fluctuations at the high electron temperatures following femtosecond laser excitation. 

\begin{figure}
	\centering
	\includegraphics[width=12cm]{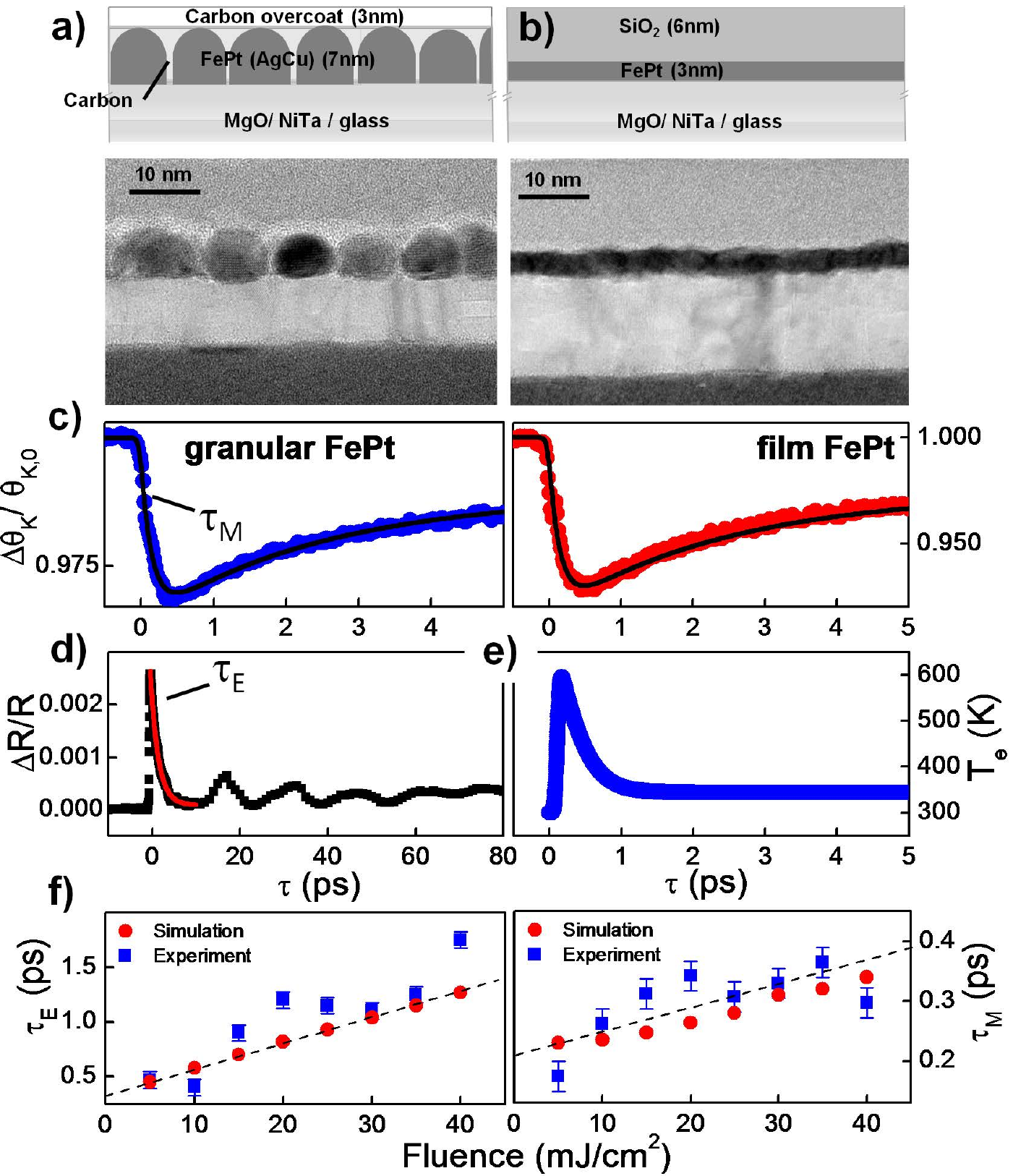}
\caption{\label{fig:tau_E} Sample characteristics: schematics and sample structure of the granular FePt recording media a) and thin film b) as measured by transmission electron microscopy (TEM). c) Ultrafast magnetization dynamics for both cases after femtosecond laser excitation. Solid line: analytical three temperature model to obtain $\tau_M$. Both can be described with sets of identical parameters. d) The reflectivity dynamics from which the exponential decay $\tau_E$ and e) the electron temperature $T_e$ are obtained. f) The relaxation time $\tau_E$ for the electron temperature and $\tau_M$ for the ultrafast demagnetization is given below for a set of pump fluences. }
\end{figure}

Even though CoPt$_3$ was among the first thin film systems investigated\cite{Beaurepaire CoPt3} since the discovery of ultrafast demagnetization in 1996 by Beaurepaire et al.\cite{Beaurepaire}, most investigations were centered on samples with in-plane anisotropy, so that little is known about the behavior of materials with perpendicular anisotropy. A notable exception is ferrimagnetic CoFeGd, which was studied in all-optical ultrafast switching triggered by a single laser pulse\cite{Stanciu,Ostler}. The modeling of this mechanism involves the complex interaction of the two spin subsystems\cite{Radu} and is still under debate. To enable progress in high-speed and high-capacity magnetic storage devices, a fundamental understanding of the ultrafast demagnetization dynamics in these materials is required.

Recent work of Koopmans et al.\cite{Koopmans2010} suggests the classification of materials as "fast" (or type I) and "slow" (or type II) based on the ratio $T_C/\mu_{at}$, where $\mu_{at}$ is the magnetic momentum per atom and $T_C$ is the Curie temperature. In both cases, there is an initial sub-picosecond fast demagnetization. However, in the first case the fast femtosecond demagnetization is followed by a magnetization recovery (as in Ni \cite{Atxitia}), while in the latter a second slower demagnetization takes place. The recovery occurs on the timescale on the order of 50 ps and more (as in Gd\cite{Sultan}). According to this classification FePt should be regarded as a fast magnetic material. However, more recently it has been shown that in Ni both behaviors can be observed, depending on the amount of deposited energy\cite{Koopmans2012}. Thus, the question of whether FePt can behave as  "fast" or "slow" under specific laser excitation is an open question. In addition, for thin films and granular media the contributions of spin currents to the ultrafast demagnetization dynamics cannot be neglected \cite{Battiato,Rudolf,Melnikov}. In this work, we use isolating substrates and cap layers to minimize these effects. The ultrafast demagnetization dynamics of a FePt continuous film sample and a high anistropy granular recoding medium is investigated. A transition from type I to type II is found, triggered by the laser fluence. We pinpoint the electron temperature reached by the laser heating as the underlying mechanism in FePt.

We have studied a 3~nm-thick continuous FePt thin layer ($H_C=200$~mT) and 7 nm-thick AgCuFePt-C granular recording media ($H_C=2.4$~T), shown in Fig. \ref{fig:tau_E} a) and b). In the granular media the carbon intercalated the magnetic 5-8~nm grains to separate them magnetically. The continuous FePt film is covered with a 6 nm-thick SiO$_2$ film to ensure that it remains smooth and continuous during the fabrication process\cite{McCallum}. The granular film has a 3 nm protective carbon overcoat, a small amount of Ag to reduce the FePt Ll$_0$ ordering temperature, and a small amount of Cu to lower the  $T_C$. The whole structure from bottom to top is glass/NiTa/MgO/FePt/SiO$_2$ (glass/NiTa/MgO/AgCuFePt-C/carbon overcoat). The continuos film sample has a Curie temperature of 650 K, a saturation magnetization $M_s=1070$ emu/cm$^3$, a maximum anisotropy constant $K_{u,max}=2.2\cdot10^7$ erg/cm$^3$, and an average anisotropy constant $K_{u,av}=1.4\cdot10^7$ erg/cm$^3$. The presented data are obtained by fluence-dependent all-optical pump-probe experiments. In Fig. \ref{fig:tau_E} c) the time-resolved magneto-optical Kerr effect ($\Delta\theta_{K}/\theta_{K,0}$) is shown for the granular FePt recording media as well as the data for the continuous FePt layer. Besides the absolute scale, both can be described using identical parameters of the analytical solution of a rate equation model shown as a continuous line: the microscopic mechanisms on the nanometer lenght scale dominate the dynamics on a femtosecond time scale. This leads to identical spectra for the  continous film and granular media. Spin currents transmitted through the carbon interfaces have no influence on the dynamics on the picosecond scale here. Moreover for the granular structure in case of the recording media, the larger $H_C$ and different $K_{u}$ does not alter the magnetization dynamics. The energy scale of the magnetic anisotropy $K_{u}< 1$ meV is too small to affect the dynamics on the timescale related to the energy scale of the of exchange interaction.

\begin{figure}
	\centering
	\includegraphics[width=12cm]{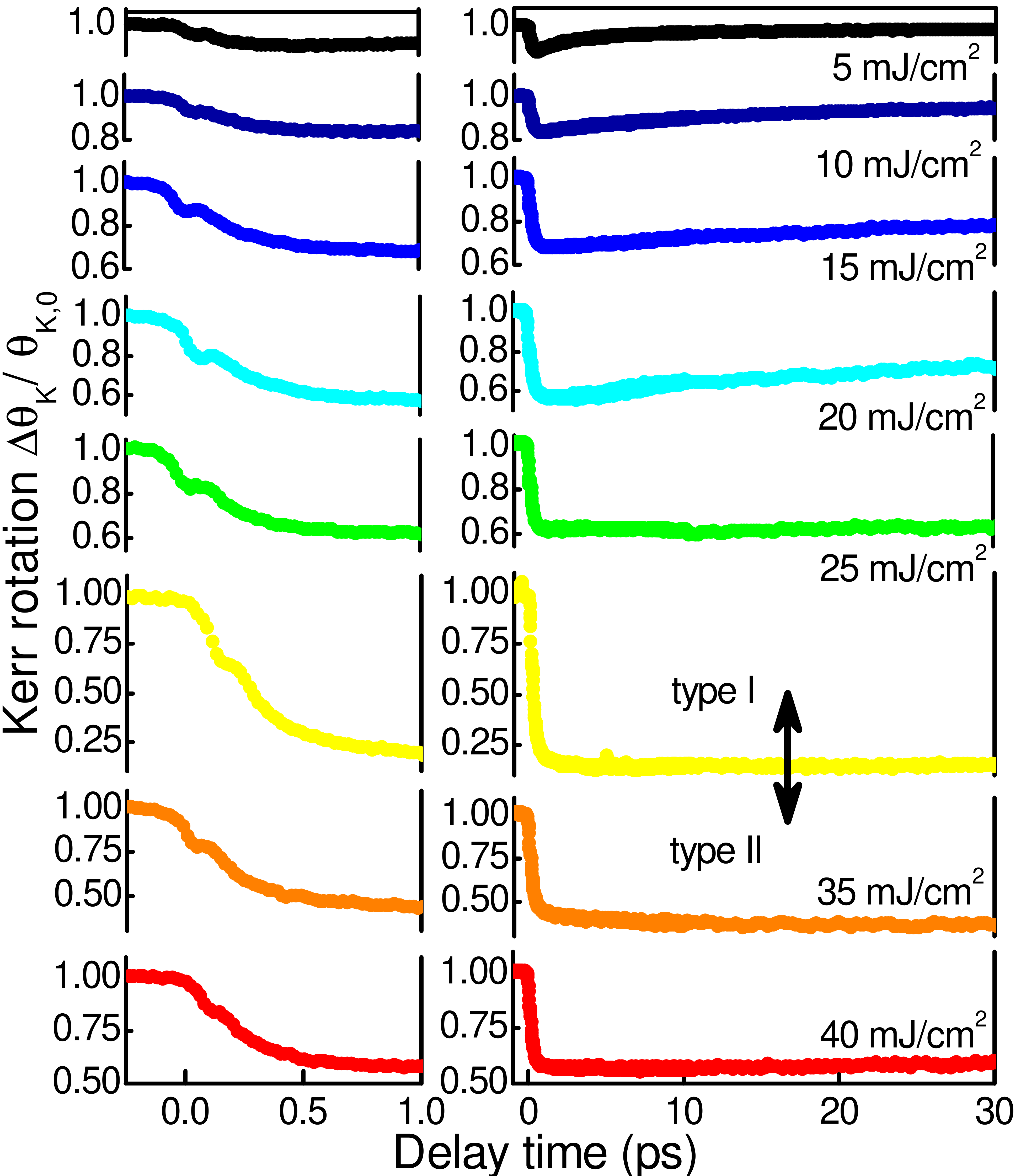}
\caption{\label{fig:Dynamics}
Ultrafast demagnetization dynamics of FePt: spin dynamics measured by the Kerr set-up at increasing laser fluence in steps of 5 mJ/cm$^2$ from 5 mJ/cm$^{2}$ (upper curve) to 40 mJ/cm$^{2}$ (lower curve). The detail on the femtosecond timescale is shown with expanded scale on the left side.
}
\end{figure}

We extract the microscopic parameters of the ultrafast magnetization dynamics of FePt in polar geometry using moderate B-fields for switching the magnetization. In the experiment, the fluence of the pump beam is varied from 5 to 40 mJ/cm$^2$ in steps of 5 mJ/cm$^2$. The magneto-optical Kerr rotation of the probe beam is measured\cite{Walowski2008} and its time delay relative to the pump beam is varied. Similarly, the time resolved reflectivity is detemined. The decay of the reflectivity signal is fitted to a simple exponential function before characteristic stress waves set in (Fig. \ref{fig:tau_E} d)), according to \mbox{$R (t) \sim \exp{(-t/\tau_E)}$}. The results are presented in Fig. \ref{fig:tau_E} f) showing the evolution of the characteristic timescale $\tau_E$ that represents the relaxation time of the electron temperature. Experimentally, the value for $\tau_E\sim$ 1 ps is a typical value expected for transition metals and consistent with previous results\cite{Atxitia2011}.

The Kerr rotation is extracted from that of opposite external field direction in order to remove all non-magnetic and thus symmetric contributions\cite{Kampfrath}. To get the absolute degree of demagnetization, the Kerr signal is scaled to hysteresis measurements at two states of reference; one at negative delay ($\theta_{K}\sim M_{z,0}$) and the other at a time delay that shows the lowest magnetization ($\Delta \theta_{K,min}\sim\Delta M_{z,min}$). The pulse shape is assumed to be Gaussian and has a full width at half maximum of $\tau_{p}=$ 40 fs. The repetition rate of the laser system is 250 kHz and thus every 4 $\mu$s a pulse excites the sample. A magnetic field of $\mu_0H=200$ mT perpendicular to the film plane, parallel to the easy axis, is applied. The magnetization dynamics $\Delta M_{z}$ (Kerr rotation angle $\Delta \theta_K)$ is presented in Fig. \ref{fig:Dynamics}. The first rapid demagnetization, $\tau_M$, occurs at a timescale below 1 ps in the range 0.15-0.3 ps, increasing with the laser pump fluence. Also the remagnetization timescale slows down as a function of the incident fluence. Finally a second, slower demagnetization with the absence of recovery is found above a fluence of 30 mJ/cm$^{2}$. Thus, a transition from type I to type II as classified in Ref.\cite{Koopmans2010} is observed for FePt.

\begin{figure}
	\centering
	\includegraphics[width=12cm]{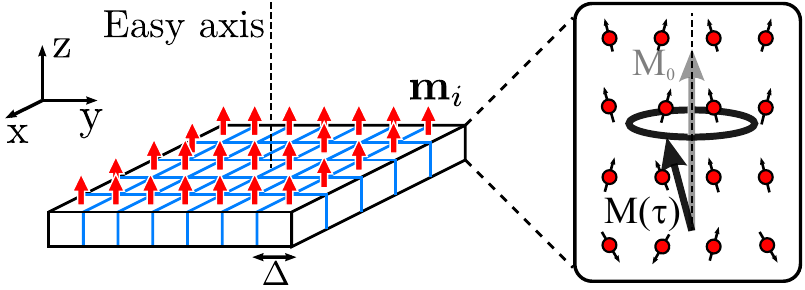}
\caption{\label{fig:scheme} Micromagnetic Landau-Lifshitz-Bloch model: the magnetization is described by the average over 900 thermal macrospins with a lateral cubic discretization of $\Delta=3$ nm with periodic boundary conditions.
Each cell represents thermodynamic average over atomistic  magnetic moments (shown schematically on the right side). Within each cell,  the precession and relaxation  along and around the quantization direction  (the longitudinal relaxation $M(\tau)$ and the rotation of the individual macrospin on the nanometer sized element) are taken into account. }
\end{figure}

To understand the behavior, we model the ultrafast magnetization dynamics under external laser excitation by a thermal process of electronic origin \cite{Atxitia}.  The model considers that within a timescale of the order of 10 fs the electrons are thermalized and can be described by a quasi-equilibrium electron temperature $T_e(t)$, coupled to the phonon and spin systems. A multi-macrospin model is used with cubic discretization elements with a lateral size of $\Delta=3$ nm (and thus a volume of $V=\Delta^3$) as illustrated in Fig. \ref{fig:scheme}. The thermal dynamics of the spin system within each cell is described macroscopically within the Landau-Lifshitz-Bloch (LLB) micromagnetic formalism\cite{Atxitia2007}. For the present simulation, a system of $30\times30\times1$ macrospins with periodic boundary conditions in $x$ and $y$ directions is used. Then every single macrospin $\textbf{m}_{i}=\textbf{M}_{i}/M_{e}(0)$  is described using the LLB equation for a finite spin $S$ that reads \cite{Garanin, Garanin2004}:

\begin{figure}
	\centering
	\includegraphics[width=12cm]{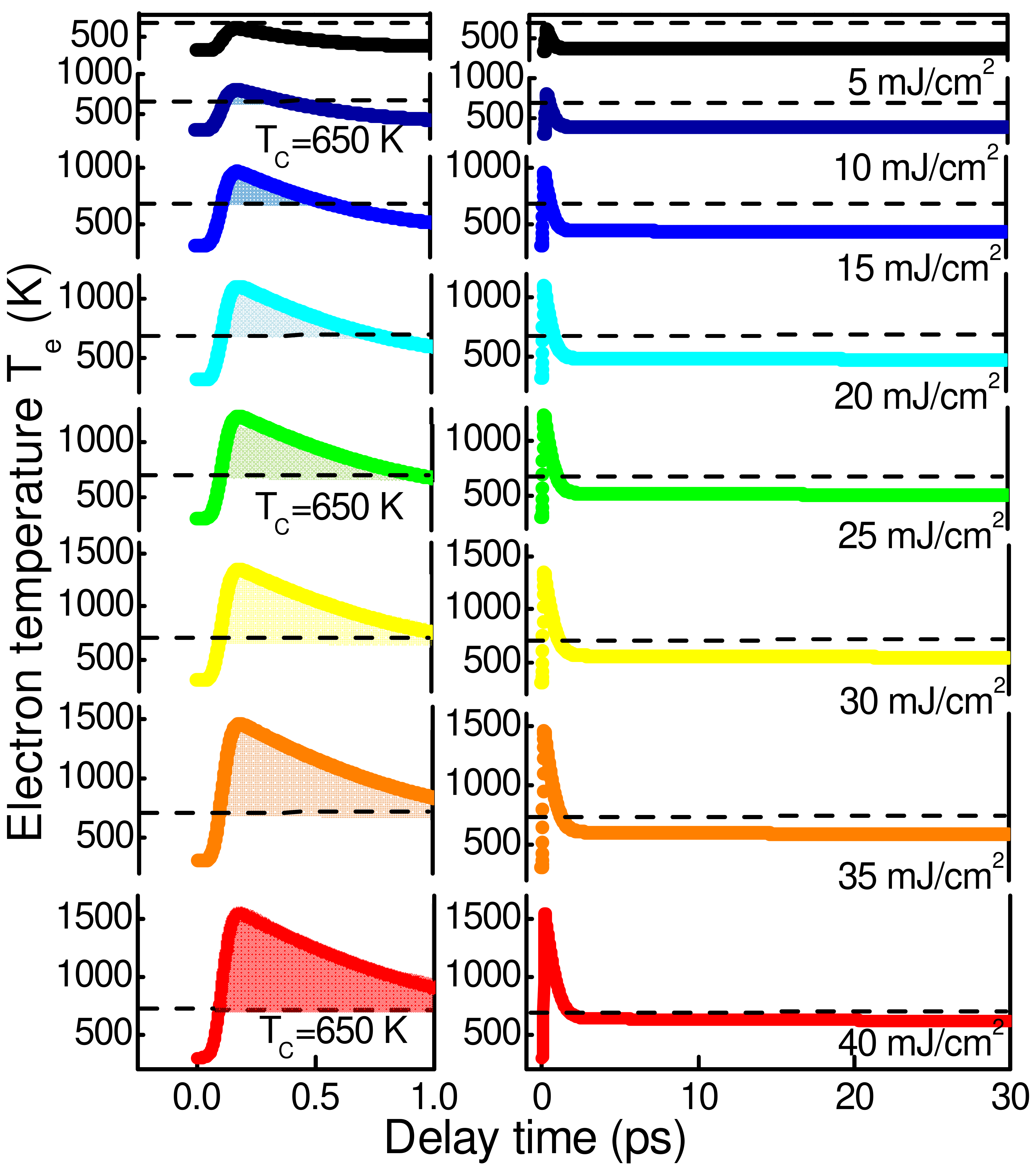}
	\caption{\label{fig:etemp} Simulation of the electron temperature $T_e$, shown as a function of the laser pump fluence (from 5 mJ/cm$^{2}$ (upper curve) to 40 mJ/cm$^{2}$ (lower curve), in steps of 5 mJ/cm$^{2}$. The 2T model is based on the set of parameters presented in Table I. The parameters are extracted from the reflectivity dynamics (Fig. \ref{fig:tau_E}). Within the shaded area marked in the left panel, the electron temperature exceeds the Curie temperature.}
\end{figure}

\begin{eqnarray}
	 \frac{d\textbf{m}_i}{dt}=&&\gamma[\textbf{m}_i\times\textbf{H}^i_{eff}]-
\frac{\gamma\tilde\alpha_\parallel}{m_i^2}(\textbf{m}_i\cdot\textbf{H}_{eff}^i)\textbf{m}_i\\ &&+\frac{\gamma\tilde\alpha_\perp}{m_i^2}[\textbf{m}_i\times[\textbf{m}_i\times
(\textbf{H}_{eff}^i+\vec{\zeta}_{i,\perp})]]+\vec{\zeta}_{i,ad}\nonumber.
\end{eqnarray}

$\textbf{M}_i$ is the spin polarization (thermal average of atomistic spins over the volume $V$ at temperature $T$), $M_e(T)$ is its equilibrium value and $M_e(0)$ is the maximum spin polarization at $T=0^o$ K and $M_e(300^o K)=M_s$.
The value of $M_e(T)$ is evaluated in the mean-field approximation (MFA) via the Brillouin function. For FePt, it has been shown \cite{Lyberatos} that the best fit for the temperature-dependent experimentally measured magnetization is obtained with the spin value $S=3/2$. The effective field $\textbf{H}^{i}_{eff}=\textbf{H}+\textbf{H}_{i,A}+\textbf{H}_{i,EX}+\textbf{H}_{J}$ is comprised of applied, anisotropy, micromagnetic exchange, and internal exchange fields.
The micromagnetic anisotropy  field $\textbf{H}_{i,A}=-\frac{1}{\tilde{\chi}_{\bot}}\left(m_{i,x}\textbf{e}_{x}+m_{i,y}\textbf{e}_{y}\right)$ is determined by the perpendicular susceptibility \mbox{$\tilde{\chi}_{\bot}=\partial m/\partial H_{\bot}$}. Experimental \cite{Okamoto} and theoretical \cite{Mryasov} results report that in FePt the anisotropy scales with magnetization as \mbox{$K \sim m^{2.1}$}. Thus, we use $\tilde{\chi}_{\bot}=M_{s}(0)/2K(0)m^{0.1}$. The micromagnetic exchange field is defined as \cite{Kazantseva}
\begin{equation}
\textbf{H}_{i,EX}=\frac{A_i(T)}{m_{i}^{2}}\frac{2}{M_{s}\bigtriangleup^{2}}\sum_{j}\left(\textbf{m}_{j}-\textbf{m}_{i}\right)
\end{equation}
where $j$ goes over neighboring elements and $A(T)$ is the micromagnetic exchange stiffness, for FePt it has been shown \cite{Atxitia2010} to scale with the magnetization  as  $A(T)\sim m^{1.76}$,  where $A(0)=2.2\cdot10^{-6}$ erg/cm, thus, we use $A_i(T)=A(0) m_i^{1.76}$.
The internal  exchange field $\textbf{H}_{J}$ results from the thermal average of atomic spins, comprising a sufficiently large discretization volume $V$ in the MFA. At low temperatures, it is responsible for keeping the magnetization magnitude constant. It is described by the following expression:
\begin{equation}
\textbf{H}_{J}=\begin{cases}
\frac{1}{2\tilde{\chi}_{\parallel}}\left(1-\frac{m_{i}^{2}}{m_{e}^{2}}\right)\textbf{m}_{i}  & T\lesssim T_{C}\\
-\frac{1}{\tilde{\chi}_{,\parallel}}\left(1+\frac{3T_{C}}{5\left(T-T_{C}\right) }m_{i}^{2}\right)\textbf{m}_{i}  & T\gtrsim T_{C}.
\end{cases}
\end{equation}
Here $\tilde\chi_{\parallel}=\partial m/\partial H_{||}$  represents the longitudinal susceptibility, evaluated in the MFA as
\begin{equation}
\tilde{\chi}_{\parallel}
 =  \frac{\mu_{0}\beta B'_{S}(\xi_{e})}{1-\beta S^{2}J_{0}B'_{S}(\xi_{e})}\quad;\quad\xi_{e}=\frac{3ST_{C}m_{e}}{(S+1)T}\label{eq:suscp_parallel}
\end{equation}
 where $B'()$  stands for the derivative of the Brillouin function. The relationship between the internal exchange parameter $J_{0}$ (also related to $A(T=0$ K), see Ref.\cite{Atxitia2010})  and $T_{C}$ is given by $T_{C}=S(S+1)J_{0}/3k_{B}$ where $k_B$ is the Boltzmann's constant. The stochastic fields $\boldsymbol{\zeta}_{i,\bot}$ and $\boldsymbol{\zeta}_{i,ad}$ are given by \cite{Evans2012}
\begin{eqnarray}
\left\langle \zeta^{k}_{i,\bot}(0)\zeta^{l}_{j,\bot}(t)\right\rangle
& = &\frac{2\vert\gamma\vert k_{B}T\left(\tilde{\alpha}_{\bot} -\tilde{\alpha}_{\parallel}\right)}{M_{s}V\tilde{\alpha}_{\bot}^{2}}\delta_{ij}\delta_{kl}\delta(t)\\
\left\langle \zeta^{k}_{i,ad}(0)\zeta^{l}_{j,ad}(t)\right\rangle
& = &\frac{2\vert\gamma\vert k_{B}T\tilde{\alpha}_{\parallel}}{M_{s}V}\delta_{ij}\delta_{kl}\delta(t)
\end{eqnarray}
where $i$ and $j$ denote the macrospin number and $k$ and $l$ denote its Cartesian components $x$, $y$, and $z$. Finally, the longitudinal and transverse relaxation parameters are \cite{Garanin2004}
\begin{equation}
\label{eq:relax_parallel}
\tilde{\alpha}_{\parallel}
 = \lambda\frac{2T}{3T_{c}}\frac{2q_{s}}{\sinh(2q_{s})} \quad
\tilde{\alpha}_{\bot}=\lambda\left[\frac{\tanh(q_{s})}{q_{s}}-\frac{T}{3T_{C}}\right]
\end{equation}
where $\lambda$ is the microscopic relaxation parameter that couples the spin dynamics to the electron temperature, defined by the microscopic spin scattering rate, and $q_{s}=3T_{C}m_{e}/[2(S+1)T]$.

The magnetization dynamics is coupled to the electron temperature $T_e(t)$. In turn, the electron temperature within the two temperature model (2T)  is coupled to the lattice temperature $T_{ph}(t)$ via rate equations:
\begin{eqnarray}
	C_e\frac{\text{d}T_e}{\text{d}t}&&=-G_{e-ph}(T_e-T_{ph})+P(t)-C_e\frac{(T_e-T_{room})}{\tau_{ph}}\nonumber\\
  C_{ph}\frac{\text{d}T_{ph}}{\text{d}t}&&=G_{e-ph}(T_e-T_{ph}).\\\nonumber
\end{eqnarray}

Here $C_e$ and $C_{ph}$ denote the specific heat of the electrons and the lattice, respectively,  $G_{e-ph}$ is the coupling constant determining the energy exchange between the electron and lattice systems,  and $\tau_{ph}$ is the heat diffusion time to the substrate. For $C_e$, the free electron approximation is used resulting in $C_e=\gamma_eT_e$. $C_{ph}$ is set constant, since FePt has a Debye temperature well below the room temperature $T_{room}$. The laser absorbed power is defined by $P(t)=I_0F\exp{[-(t/\tau_p)^2]}$ proportional to the laser pump fluence $F$.  The time resolved reflectivity reveals that the change of electron temperature depends linearly on the change in reflectivity\cite{Caffrey}. Thus, we assume that the lacking parameters of the 2T model can be extracted from the measured reflectivity at the lowest pump fluence $F=5$ mJ/cm$^2$.  The long-term diffusion timescale $\tau_{ph}$ is obtained from the long-term magnetization behavior. The proportionality constant $I_0$ is estimated by fitting of the experimental demagnetization value at $30$ ps, and the coupling-to-the bath parameter $\lambda$ via matching the maximum demagnetization value. Note that the value  obtained for $\lambda$ is similar to those used previously for the simulations of FePt \cite{Kazantseva2008} and corresponds to an enhanced spin scattering rate at high temperatures.

The determination of the material specific constants such as $\gamma_e$ and $G_{e-ph}$ is crucial for a proper simulation of the demagnetization process. Two approaches were performed, resulting in a high and low electron temperature (see Supplementary Materials). First, $G_{e-ph}=1.5\cdot10^{17}$ W/m$^3$K is assumed and thus in the range of Cu, Mo, and Pt \cite{Hohlfeld}. Then, the fitting to the reflectivity relaxation rate, measured at $F=5$ mJ/cm$^2$,  gives $\gamma_e=110$ J/m$^3$K$^2$. This value is of the order reported for Au and Cu \cite{Hohlfeld} but is much smaller than the corresponding value for Ni\cite{Atxitia} and hence produces a high electron temperature. The final set of parameters is presented in Table I. The corresponding simulated electron temperature is shown in Fig. \ref{fig:etemp}. The electron temperature decay fitted to the same exponential decay function as in the experiment and the relaxation time $\tau_E$ shows a reasonable agreement with the experiment for all fluences (Fig. \ref{fig:tau_E} f)).
In the second case, a coupling constant $G_{e-ph}=1.8\cdot10^{18}$ W/m$^3$K, similar to Ni \cite{Atxitia} is assumed which gives $\gamma_e=1700$ J/m$^3$K$^2$, more proper to transition metals, and thus about one order of magnitude larger than in the first case. As a consequence, a lower electron temperature (with the maximum value up to $1000$ K) is reached.

\begin{figure}
	\centering
	\includegraphics[width=12cm]{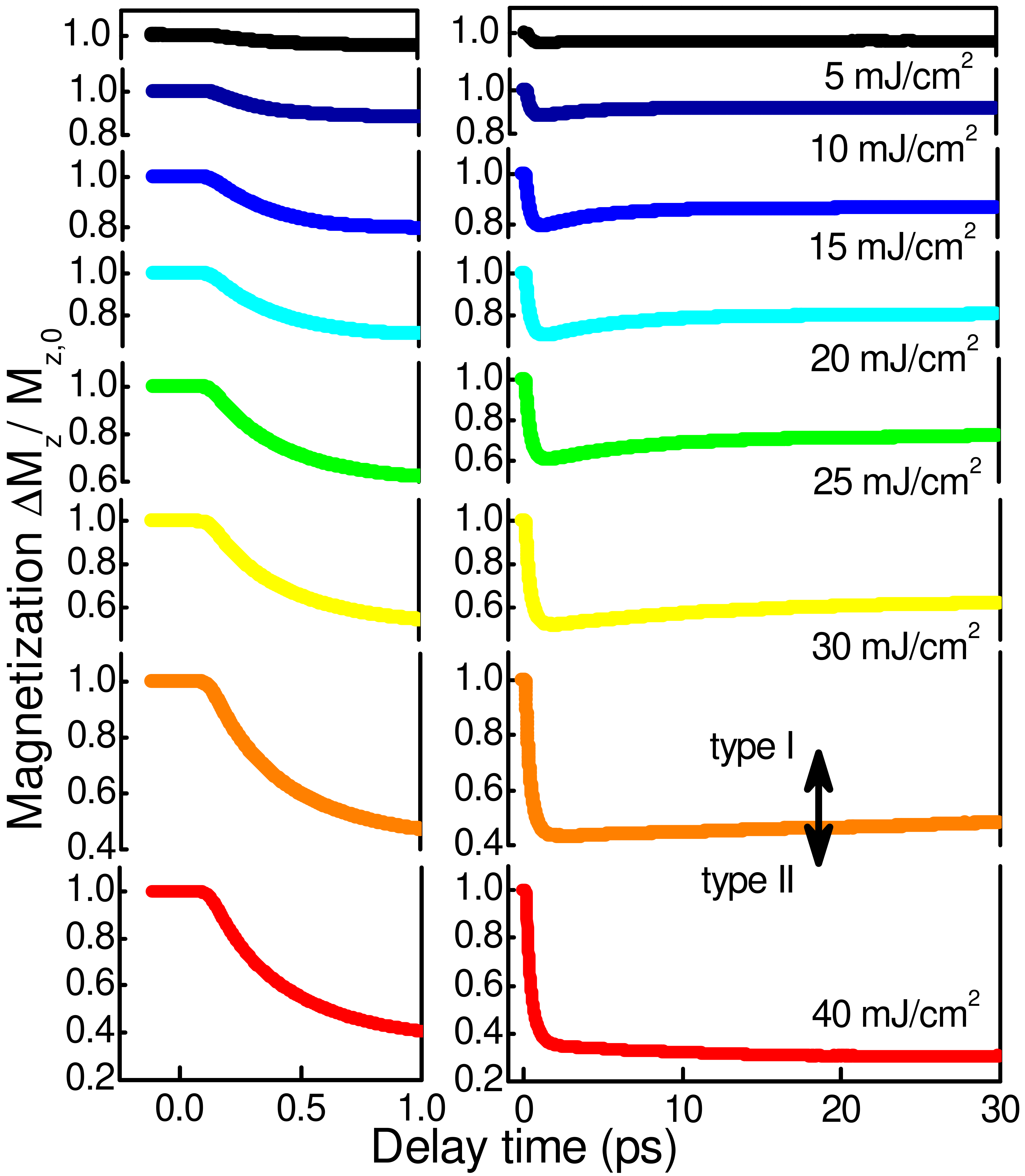}
	\caption{\label{fig:theory}Ultrafast demagnetization dynamics obtained by integration of the LLB micromagnetic model coupled to the 2T model (Fig. \ref{fig:etemp}). The model data are shown with increased laser fluence from top to bottom in steps of 5 mJ/cm$^2$ from 5 mJ/cm$^{2}$ (upper curve) to 40 mJ/cm$^{2}$ (lower curve). The detailed view on the femtosecond timescale is shown with expanded scale on the left side.}
\end{figure}

\setlength{\extrarowheight}{2pt}
\begin{table}
	\caption{\label{tab:hightemp}Overview of all constants relevant for the simulation in the case of high electron temperature.}
\begin{ruledtabular}
\begin{tabular}{ccccccc}
	$\gamma_e$	&$\lambda$	&$I_0$		&$G_{el-ph}$		&$C_{ph}$		&$\tau_{ph}$	&$S$	\\
     (J/m$^3$K$^2$)	&(a.u.)		&(s$^{-1}$)	&(W/m$^3$K)		&(Jm$^{-3}$K$^{-3}$)	&(ps)		&($\hbar$)\\	
\hline
110		&0.1		&5.0$\cdot10^{16}$      &1.5$\cdot10^{17}$    &3.7$\cdot10^{5}$	&340		&3/2		\\
\end{tabular}
\end{ruledtabular}
\end{table}

The results for the integration of the set of the LLB equations, coupled to the 2T model, with the set of parameters from Table I are presented in Fig. \ref{fig:theory} for all fluences. As in the experiment, the simulations show a transition between type I and type II behavior. The agreement is found in the demagnetization values $\Delta M/M(300K)=0.05-0.7$ as well as in the demagnetization times $\tau_M= 0.2-0.3$ ps. The theory shows a linear increase of both values. Contrary to this, the integration of the LLB equation coupled to the 2T model with the second set of parameters (lower electon temperature) does not produce a transition between type I and type II behavior  and the sub-ps demagnetization is always followed by the remagnetization within several ps (see Supplementary Materials).  We conclude that the transition is defined by a critical temperature that the electrons have reached. This is illustrated by the shaded regions in Fig. \ref{fig:etemp}. Only if the electron temperature $T_{e}$  stays near the Curie temperature $T_C$ for several picoseconds this transition is found. From the theory of phase transitions, we know that at such temperatures the dynamics will be characterized by an increased dominance of magnetization fluctuations. Particularly, the divergence of correlation lengths leads to slowing down of correlation times \cite{Fischer, ChubykaloFesenko}. We find that the characteristics of type II materials are related to a non-deterministic fractioning into dynamic spin excitations. In the experiment (Fig. \ref{fig:Dynamics}) the relative decay is deviating for the two highest laser fluences, in comparison to the model (Fig. \ref{fig:theory}). This discrepancy is related to the decrease of the magnetization at negative delay due to the accumulation of the high pump power, which is not taken into acount in the LLB model.

In summary, by means of the time-resolved Kerr magnetometry we have investigated ultrafast magnetization dynamics in FePt thin films with perpendicular anisotropy. Our results indicate that the amount of the absorbed energy plays a crucial role in the character of the ultrafast demagnetization in FePt. The measurements reveal a transition from type I to type II behavior. Our experimental results are modeled in terms of the micromagnetic LLB model, coupled to the 2T model. Within this framework, we find that transition to type II behavior is a consequence of high electron temperature. We identify that at large pump fluences the resulting electron temperature remains close to the Curie temperature and is leading to critical magnetization fluctuations \cite{ChubykaloFesenko} responsible for this transition. This non-deterministic spin dynamics is responsible for a speed limtitation of the magnetic response to the laser pulse. Note that this is defined not only by the laser fluence, but also by the nature of the FePt's density of states at the Fermi level defining the increase in electron temperature. Our results open possibilities for ultrafast control of the demagnetization in FePt, the most promising candidate for future magnetic recording applications. Importantly, we have shown that we are able to manipulate the degree of demagnetization and its ultrafast rates. We propose that for efficient writing the degree of heating and its speed have to be balanced by varying the amount of the energy deposited.

Acknowledgments: \noindent we thank A. McCallum for providing a sample and TEM image,  O. Mosendz for a TEM image and U. Atxitia for his help with the theory. Research at G\"ottingen University was supported by German Research Foundation (DFG) through SFB~602, MU 1780/ 6-1 Photo-Magnonics and SPP~1538 SpinCaT. Research in Madrid was supported by the European Community's Seventh Framework Programme under grant agreement NNP3-SL-2012-281043 (FEMTOSPIN) and the Spanish Ministry of Science and Innovation under the grant FIS2010-20979-C02-02.


\begin{references}
\bibitem{Moser} A. Moser and D. Weller, in \emph{The physics of high density magnetic recording} (Springer, New York, 2001), p. 145.
\bibitem{Stipe} B. C. Stipe et al., Nature Photon. 4, 484 (2010).
\bibitem{Back} I. Tudosa, C. Stamm, A. B. Kashuba, F. King, H. C. Siegmann, J. St\"ohr, G. Ju, B. Lu, and D. Weller, Nature 428, 831 (2004). C. Back, R. Allenspach, W. Weber, S.S. Parkin, D. Weller, E. L. Garwin, H. C. Siegmann, Science 285,864 (1999).
\bibitem{Stanciu} C. D. Stanciu, F. Hansteen, A. V. Kimel, A. Kirilyuk, A. Tsukamoto, A. Itoh, and T. Rasing, Phys. Rev. Lett. \textbf{99}, 047601 (2007).
\bibitem{Beaurepaire CoPt3}E. Beaurepaire, M. Maret, V. Halté, J.-C. Merle, A. Daunois, and J.-Y. Bigot, Phys. Rev. B \textbf{58}, 12134 (1998).
\bibitem{Beaurepaire} E. Beaurepaire, J.-C. Merle, A. Daunois, and J.-Y. Bigot,  Phys. Rev. Lett. \textbf{76}, 4250 (1996).
\bibitem{Ostler} T. A. Ostler, et al., Nature Comm. 3,  \textbf{666} (2012).
\bibitem{Radu} I. Radu, K. Vahaplar, C. Stamm, T. Kachel, N. Pontius et al., Nature \textbf{472}, 205 (2011).
\bibitem{Koopmans2010} B. Koopmans, G. Malinowski, F. Dalla Longa, D. Steiauf, M. F\"{a}hnle, T. Roth, M. Cinchetti and M. Aeschlimann, Nature Mater. \textbf{9}, 259265 (2010).
\bibitem{Battiato}M. Battiato, K. Carva, P. M. Oppeneer, Phys. Rev. Lett. \textbf{105}, 027203 (2010).
\bibitem{Rudolf}D. Rudolf, C. La-O-Vorakiat, M. Battiato, R. Adam, J.M. Shaw et al., Nature Comm. \textbf{3}, 1037 (2012).
\bibitem{Melnikov}A. Melnikov, I. Razdolski, T. O. Wehling, E. Th. Papaioannou, V. Roddatis et al., Phys. Rev. Lett. \textbf{107}, 076601 (2011).
\bibitem{Atxitia} U. Atxitia, O. Chubykalo-Fesenko, J. Walowski, A. Mann and M. M\"unzenberg, Phys. Rev. B  \textbf{81}, 174401 (2010).
\bibitem{Sultan} M. Sultan, U. Atxitia, A. Melnikov, O. Chubykalo-Fesenko and U. Bovensiepen, Phys. Rev. B \textbf{85}, 184407 (2012).
\bibitem{Koopmans2012}T. Roth, A. J. Schellekens, S. Alebrand, O. Schmitt, D. Steil, B. Koopmans, M. Cinchetti, M. Aeschlimann, Phys. Rev. X \textbf{2}, 0221006 (2012).
\bibitem{McCallum} A. T. McCallum, D. Kercher, J. Lille, D. Weller, and O. Hellwig,  Appl. Phys. Lett. \textbf{101}, 092402 (2012).
\bibitem{Kampfrath} T. Kampfrath, R. G. Ulbrich, F. Leuenberger, M. M\"unzenberg, B. Sass, and W. Felsch, Phys. Rev. B \textbf{65}, 104429 (2002).
\bibitem{Atxitia2011} U. Atxitia and O. Chubykalo-Fesenko, Phys. Rev. B \textbf{84}, 144414 (2011).
\bibitem{Walowski2008} J. Walowski, M. Djordjevic Kaufmann, B. Lenk, C. Hamann, J. McCord and M. M\"unzenberg, J. Phys. D: Appl. Phys. \textbf{41}, 164016 (2008).
\bibitem{Atxitia2007} U. Atxitia, O. Chubykalo-Fesenko, N. Kazantseva, D. Hinzke, U. Nowak, and R. W. Chantrell, Appl. Phys. Lett. \textbf{91}, 232507 (2007).
\bibitem{Garanin}D. A. Garanin, Physica A \textbf{172}, 470 (1991)
\bibitem{Garanin2004} D. Garanin and O. Chubykalo-Fesenko, Phys. Rev. B \textbf{70}, 212409 (2004).
\bibitem{Lyberatos} A. Lyberatos and K. Yu Guslienko J.Appl. Phys. \textbf{94}, 1119 (2003).
\bibitem{Okamoto} S. Okamoto, N. Kikuchi, O. Kitakami, M. Miyazaki, Y. Shimada and K. Fukamichi, Phys. Rev. B \textbf{66}, 024413 (2002).
\bibitem{Mryasov}O. N. Mryasov, U. Nowak, K. Y. Guslienko, and R. W. Chantrell, Europhys. Lett. \textbf{69}, 805 (2005).
\bibitem{Kazantseva} N. Kazantseva, D. Hinzke, U. Nowak, R. W. Chantrell, U. Atxitia and O. Chubykalo-Fesenko, Phys. Rev. B \textbf{77}, 184428 (2008).
\bibitem{Atxitia2010} U. Atxitia, D. Hinzke, O. Chubykalo-Fesenko,  U. Nowak, H. Kachkachi, O. N. Mryasov, R. F. Evans and R. W. Chantrell, Phys. Rev.B \textbf{82}, 13440 (2010).
 \bibitem{Evans2012} R. F. L. Evans, D. Hinzke, U. Atxitia, U. Nowak, R. W. Chantrell and O. Chubykalo-Fesenko, Phys. Rev. B \textbf{85}, 014433 (2012)
 \bibitem{Caffrey}A. P. Caffrey, P. E. Hopkins, J. M. Klopf and P. M. Norris, in \emph{Thin Film Non-Noble Transition Metal Thermophysical Properties} (Taylor \& Francis Inc., 2005), p. 365-377.
 \bibitem{Kazantseva2008} N. Kazantseva, U. Nowak, R. W. Chantrell, J. Holhfeld and A. Rebei, Europhys. Lett. \textbf{81}, 27004 (2008)
 \bibitem{Hohlfeld} J. Hohlfeld and S. Wellershoff, Chem. Phys. \textbf{251}, 237 (2000)
 \bibitem{Fischer} D. Fisher, Phys. Rev. Lett. \textbf {56}, 416 (1986)
\bibitem{ChubykaloFesenko} O. Chubykalo-Fesenko, U. Nowak, R. W. Chantrell and D. Garanin, Phys. Rev. B \textbf{74}, 094436 (2006).

\end{references}

\end{document}